\documentclass[12pt]{article}

\usepackage{times}
\usepackage{graphics}
\usepackage{amssymb}
\usepackage{amsmath}

\renewcommand{\thefootnote}{\fnsymbol{footnote}}

\newcommand{\nc}{\newcommand}
\newcommand{\rnc}{\renewcommand}

\rnc{\baselinestretch}{1.24}  
\rnc{\arraystretch}{1.24}     

\nc{\be}{\begin{eqnarray}}
\nc{\ee}{\end{eqnarray}}
\nc{\xx}{\nonumber\\}

\nc{\eq}[1]{(\ref{#1})}
\nc{\newcaption}[1]{\centerline{\parbox{6in}{\caption{#1}}}}

\nc{\fig}[3]{
\begin{figure}
\centerline{\epsfxsize=#1\epsfbox{#2.eps}}
\newcaption{#3. \label{#2}}
\end{figure}
}

\def\CT{{\cal T}}

\def\a{\alpha}
\def\b{\beta}
\def\g{\gamma}
\def\d{\delta}

\def\l{\lambda}
\def\m{\mu}
\def\n{\nu}

\def\s{\sigma}

\def\w{\omega}

\def\G{\Gamma}
\def\D{\Delta}

\def\L{\Lambda}

\def\half{\frac{1}{2}}
\def\imply{\Longrightarrow}

\def\goto{\rightarrow}

\def\del{\nabla}

\def\p{\partial}

\def\Tr{{\rm Tr}}

\def\det{{\rm det}}

\def\bp{\bar{\partial}}
\def\bA{\bar{A}}
\def\tE{\tilde{E}}

\def\bi{\bar{i}}
\def\bj{\bar{j}}
\def\bk{\bar{k}}

\begin{document}

\begin{titlepage}

\thispagestyle{empty}

\hfill\parbox{4cm}
{hep-th/0309217 \\CERN-TH/2003-224}

\vspace{15mm}
\centerline{\Large \bf On the Geometry of Coset Models with Flux}
\vspace{10mm}

\begin{center}
Sangmin Lee
\footnote{sangmin.lee@cern.ch}
\\[7mm]
{\sl Theory Division, CERN, CH - 1211}
\\
{\sl Geneva 23,  Switzerland}
\end{center}

\vskip 40mm

\begin{abstract}
\noindent
We study the 3-form flux $H_{\m\n\l}$ associated with the 
semi-classical geometry of $G/H$ gauged WZW models. 
We derive a simple, general expression for the flux 
in an orthonormal frame and use it to explicitly verify 
conformal invariance to the leading order in $\a'$. 
For supersymmetric models, we briefly revisit the 
conditions for enhanced supersymmetry. 
We also discuss some examples of non-abelian cosets with flux.

\end{abstract}

\vspace{2cm}
\end{titlepage}


\baselineskip 7mm
\renewcommand{\thefootnote}{\arabic{footnote}}
\setcounter{footnote}{0}

\subsection*{Introduction}

WZW models and their cosets (gauged WZW) provide examples of 
string backgrounds where both the exact CFT description 
and the geometry of the target space are well-known. 
The coset space $G/H$ is obtained by the identification $g \sim h g h^{-1}$ 
$(g\in G, h\in H)$, hence its geometry is quite different from that 
of the usual left-coset ($g \sim hg$). 
The `adjoint-coset' is also required to have non-trivial 
dilaton and three-form flux ($H_{\m\n\l}$) on it  
in order to ensure conformal invariance.

For left-cosets, the invariant one-forms and structure constants 
offer a clear intuitive picture of the geometry. 
In Ref. \cite{tsey1}, analogous one-forms were introduced for 
adjoint cosets and were shown to define an orthonormal frame for the metric. 
The goal of this note is to take advantage of these one-forms 
to better understand the geometry of the adjoint coset 
with emphasis on the properties of the flux.  

We first derive a simple, general expression for the flux 
in the orthonormal frame.
\footnote{ 
Throughout this paper, we work only in the semi-classical 
($\a'/R^2 \sim 1/k \ll 1$) limit because the problem of obtaining 
the exact expression for the flux is quite involved \cite{sfet-tsey}.
} 
As a consistency check, we use it 
to verify conformal invariance to the leading order. 
We then specialize to supersymmetric cases and 
comment on the enhancement of world-sheet supersymmetry 
from $N=1$ to $N=2$ in the presence of the flux. 
Finally, we discuss the conditions for vanishing of the flux 
and two examples of non-abelian cosets with dim($G/H$) = 6. 
Our result may be useful in the study of how mirror symmetry works 
\cite{lou} (See also \cite{hen}) in an NS-NS flux background and 
the geometric aspects of D-branes in gauged WZW model \cite{mms}.

\subsection*{Setup}

We begin with a very brief review of WZW model and its cosets 
to set up our notations. Let $G$ be a compact, simple Lie group. 
The Lie algebra of $G$ is written in terms of an orthonormal basis of  
anti-Hermitian generators as
\be
[ T_{A}, T_{B} ] = f_{AB}{}^C T_C, \;\;\;\;\; \Tr(T_A T_B) = - \d_{AB} .
\ee
To describe the geometry of the group manifold, 
we introduce the standard one-forms:
\be
&& g^{-1} dg = E^A T_A, \;\;\;\; dg\; g^{-1} = \tE^A T_A \xx
&& \tE^A = C^{AB} E^B , \;\;\;\;\; C_{AB} = - \Tr(T_A g T_B g^{-1}), 
\;\;\;\;\; C C^{T} = 1.
\ee
The WZW model defined for $G$,  
\be
S_G &=& - \frac{k}{4\pi} \int d^2z \Tr ( g^{-1}\p g \cdot g^{-1} \bp g) + 
ik \G_{WZ}, 
\ee
corresponds to a sigma model on the group manifold 
with constant dilaton and the following metric and flux
\be
ds^2 = E_A E_A, \;\;\;\; H = \frac{1}{6} f_{ABC} E_A E_B E_C.
\ee
More precisely, the metric and the flux should be scaled by 
the radius square $R^2= k \psi^2 \a'/4$, 
where the integer $k$ is the level of 
WZW model and $\psi$ is the highest root of Lie($G$). 
We will suppress $R^2$ in the following unless 
its precise value becomes important. 

We will consider cosets of type $G/H$, where ${\rm rank}(H) = {\rm rank}(G)$ 
and $H$ acts on $G$ as $g \goto h g h^{-1}$.  
We use $(a,b,\cdots)$ indices for Lie($H$) and $(\a, \b, \cdots)$ indices for 
its orthogonal complement. 
The coset theory is realized as a gauged WZW theory with the following action
and gauge transformation law:
\be
S &=& S_G + S_A,  \\
S_A &=& \frac{k}{2\pi} \int d^2z \Tr ( \bA g^{-1}\p g - A \bp gg^{-1} -
\bA A + g^{-1} A g \bA ) \xx
&=& - \frac{k}{2\pi} \int d^2z (\bA_a E^a - A_a \tilde{E}^a - A^a
(\eta_{ab} - C_{ab}) \bA^b), \xx
\label{gaugerule}
g &\goto& u^{-1} g u, \;\;\;\; A_i \goto u^{-1} (A_i + \p_i) u .
\ee

\subsection*{The expression}

Since the action is quadratic in the non-propagating gauge field, 
it is easy to integrate out the gauge field and find \cite{bars2, tsey1}
\be
G_{MN} &=& G^{(0)}_{MN} + 2
(M^{-1})_{ab} E^a{}_{(M} \tilde{E}^b{}_{N)}, \\
B_{MN} &=& B^{(0)}_{MN} + 2
(M^{-1})_{ab} E^a{}_{[M} \tilde{E}^b{}_{N]}, \\
e^{-2\phi} &=& \det M, 
\ee
where $M_{ab} \equiv \d_{ab} - C_{ab}$. 
Although $G_{MN}$ and $B_{MN}$ carry $d_G = {\rm dim}(G)$ indices, 
they actually depend only on the `coset directions,' 
as can be seen from the existence of the $d_H={\rm dim}(H)$ null vectors
\be
Z_a{}^M = E_a{}^M - \tE_a{}^M = M_{ab} E_b{}^M - C_{a\b} E_{\b}{}^M 
\;\;\; \imply \;\;\; G_{MN} Z_a{}^{M} = 0.
\ee
Removal of $d_H$ degrees of freedom and gauge-invariant way 
can be made clear with the help of the one forms \cite{tsey1}
\be
H_\a = E_\a + E_a (M^{-1})_{ab} C_{b\a} \;\;\;\; (Z_a\cdot H_\a = 0) .
\ee
As shown in \cite{tsey1}, these one-forms define an orthonormal frame, 
i.e.,  
\be
ds^2 &=& H_\a H_\a .
\ee
It is natural to write down the flux also in this frame. 
A lengthy but straightforward computation 
using the basic identities, 
\be
\label{f1}
dC_{AB} &=& -C_{AD} f_{DBC} E_C , \\
\label{f2}
f_{ABC} &=& C_{AD}C_{BE}C_{CF} f_{DEF} , \\
\label{f3}
f_{ACD} f_{BCD} &=& c_G \d_{AB}, \;\;\;\; f_{acd}f_{bcd} = c_H \d_{ab}, \\
\label{f4}
f_{AB[C}f^B{}_{DE]} &=& 0, \\ 
\label{f5}
f_{ab\g} &=& 0,
\ee 
shows that the flux also takes a very simple form in this frame,   
\be
\label{expr}
H &=& \frac{1}{6} \Big\{f_{\a\b\g} 
+ 3 A_{[\a\b\g]} \Big\} 
H_\a \wedge H_\b \wedge H_\g , \xx
A_{\a\b\g} &=& f_{a\a\b} (M^{-1})_{ab} C_{b\g} .
\ee
This expression is the starting point of our discussion 
in what follows.

It is useful 
to note that the gauge transformation (\ref{gaugerule}) 
translates into a local Lorentz transformations on vielbeins $H_\a$. 
Suppose we choose a gauge slice $g_0(x)$ with a set of coordinate 
$\{x^\m\}$ ($\m = 1, \cdots, d_G - d_H$). 
Then, consider the following type of gauge transformation,
\be
\label{glorentz}
g_0(x) \goto h(f^m(x)) \; g_0(x) \; h(f^m(x))^{-1}, 
\ee
where $h(y^m)$, $(m = 1, \cdots, d_H)$ define a coordinate system on $H$. 
The functions $f^m(x)$ shift 
the gauge slice from the original one without inducing a 
coordinate change. Upon this type of gauge transformation, 
the one-forms $E_a$, $E_\a$ and $H_\a$ transform as
\be
E_a &\goto& Q_{ab}(E_b - e_c M_{cb}), \xx
\label{homo}
E_\a &\goto&  Q_{\a\b}(E_\b + e_cC_{c\b}), \\
H_a &\goto& Q_{\a\b}(x) H_\b, 
\ee
where $Q_{AB} = -\Tr(T_A h T_B h^{-1})$, and $h^{-1} dh = e_a T_a$. 
Clearly, the change of gauge slice results in a local Lorentz 
transformation on $H_\a$.

\subsection*{Conformal invariance} 

The leading order conformal invariance condition 
for a sigma model is well known to be
\be
\label{gg}
R_{MN} - \frac{1}{4} H_{MIJ} H_{N}{}^{IJ} + 2 \del_M \del_N \phi &=& 0 ,\\
\label{bb}
\del^M (e^{-2\phi} H_{MIJ} ) &=& 0 ,\\ 
\label{dil}
e^{2\phi} \del^2 ( e^{-2 \phi} ) - \frac{1}{6} H^2 &=& \L.
\ee
For WZW or coset models, the constant $\L$ on the RHS of the third equations 
equals $2 (\D d) /3\a'$, where $(\D d)$ is the deviation 
of the `dimension of the target space' (more precisely, the central charge) 
from an integer value. 

For a WZW model, it follows straight from 
$d E_A = -\half f_{ABC} E_B \wedge E_C$ that 
\be
\label{bftn}
4 R_{AB} &=& H_{ACD} H_{BCD} 
= f_{ACD} f_{BCD} = c_G \d_{AB}, \\ 
H^2 &=& \frac{c_G d_G}{R^2} = \frac{4 c_G d_G}{k \psi^2 \a'} . 
\ee
At a large $k$, the value of $H^2$ agrees with the central charge of 
the WZW model at level $k$ subtracted from its value in the 
$k\goto \infty$ limit (Recall $c =  \frac{k \psi^2 d_G}{k \psi^2 + c_G}$).
Eq. (\ref{bb}) follows from Jacobi identity for the structure constants.

For a coset space, the computation is somewhat more involved. 
As usual, the metric connection is derived from  
\be
dH_\a = -\half (f_{\a\b\g} + A_{\b\g\a} ), 
H_\b \wedge H_\g - (M^{-1})_{ab} f_{\a\b b} H_\b \wedge E_a .
\ee
The last term ensures that the spin-connection $\w_{\a\b}$ 
transform inhomogeneously under a local Lorentz transformation. 
It also produces many non-tensor terms 
in the intermediate steps of the 
computation of the curvature tensor. 
This complication can be avoided by using the gauge transformation 
(\ref{homo}) to set $E_a=0$. This can be always done at any point 
on the coset space, although care should be taken 
to include the derivatives of $E_a$, which do not vanish in general.
In this special gauge, the connection is given by 
\be
\w_{\a\b} &=& -\half (f_{\a\b\g} -A_{\a\b\g} + A_{\b\g\a} - A_{\a\g\b})H_\g
\equiv \w_{\a\b\g} H_\g ,
\ee
and the components of its derivatives that are relevant  
in computing $R_{\a\b}$ are 
\be
d (\w_{\a\b\g}) &=& \left\{ 
\half(A_{\a\b\g|\d} - A_{\b\g\a|\d} + A_{\a\g\b|\d}) 
+ \D \w_{a\b\g|\d} \right\} H_{\d}, \xx
A_{\a\b\g|\d} &=& A_{\a\b\s}(A_{\s\d\g} + f_{\s\d\g}) 
+ f_{a\b\g} (M^{-1})_{ab}C_{bc} f_{c\d\g}, \xx
2 \D \w_{\a\b[\g|\d]} &=& -(M^{-1})_{ab} f_{\a\b b}f_{a\g\d}. 
\ee
Using these results and the basic properties (\ref{f1})-(\ref{f5}), 
it is straightforward to verify the conformal invariance conditions 
(\ref{bftn}) including the precise value of $\L$.

\subsection*{${\mathbf N=2}$ Supersymmetry}

It is well-known \cite{ka-su, ka-su2} that 
supersymmetry of $N=1$ $G/H$ coset 
is enhanced to $N=2$ when 
$\CT \equiv {\rm Lie}(G)-{\rm Lie}(H)$ decomposes as $\CT = \CT_+ \oplus 
\CT_-$, where $\CT_\pm$ are complex conjugate representations of $H$ 
with $[\CT_+, \CT_+] \subset \CT_+$, $[\CT_-, \CT_-] \subset \CT_-$. 
In complex notation, closure under commutation implies that 
$f_{ijk} = 0 = f_{\bi\bj\bk}$ and $f_{ij a} = 0 = f_{\bi\bj a}$. 
It follows that the $(3,0)$ and $(0,3)$ components of the flux vanish.
This fact is in agreement with a related analysis \cite{ghr} 
of supersymmetry enhancement of sigma models in the presence of 
the flux; in Ref. \cite{ghr}, it was shown that in order for 
an $N=1$ supersymmetric sigma model to have an extra supersymmetry, 
the target space should be complex and the $(3,0)$ and $(0,3)$ 
components of the flux should vanish.  

\subsection*{Examples}

Given the formula for the flux (\ref{expr}), 
it is natural to ask what are the conditions for  
a $G/H$ coset to have non-vanishing flux. 
First, we note that the flux cannot vanish when $f_{\a\b\g}\neq 0$. 
The reason is that $f_{\a\b\g}$ and $A_{[\a\b\g]}$ are 
orthogonal to each other ($f_{\a\b\g} A_{\a\b\g} = 0$) as follows from 
(\ref{f3}) and (\ref{f5}), and therefore cannot cancel each other. 
For $N=2$ supersymmetric cosets (Kazama-Suzuki models), 
all such examples have been classified in Ref. \cite{fuchs}. 
The simplest among them is $SO(5)/SU(2)\times U(1)$ where 
$su(2)$ is embedded along a pair of long roots in $so(5)$. 

For cosets with $f_{\a\b\g}=0$, it remains to determine 
when $A_{[\a\b\g]}$ also vanishes
To our knowledge, the full answer to this question is not known. 
In the literature, all known examples with $f_{\a\b\g}=0$ and 
$A_{[\a\b\g]} \neq 0$ are abelian cosets (i.e., the subset $H$ is abelian) 
\cite{hoho, ger, rai, giqu, nawi}. 
\footnote{See \cite{pando} for an example of $(G\times G')/H$ 
coset that is rather different from the $G/H$ cosets considered here.}
Several non-abelian cosets with $f_{a\b\g} = A_{[\a\b\g]}=0$ are 
also known \cite{bars1, bars2, bars3, bars4, cre, frad, cha, lu}.
Using our formula (\ref{expr}) and 
a gauge choice similar to that of \cite{bars2}, 
we have computed the flux for the two Kazama-Suzuki models of 
dimension 6: $SU(4)/SU(3)\times U(1)$ and 
$SO(5)/SO(3)\times SO(2)$. 
It turns out that $A_{[\a\b\g]}$ vanishes for the former and
not for the latter. 
It would be interesting to develop a systematic method 
to determine whether a given coset with $f_{\a\b\g}=0$ has 
vanishing flux. Algebraic CFT description of coset models 
may turn out to be useful in that direction.

\subsection*{Acknowledgement}

I would like to thank Zheng Yin for collaboration 
at an early stage of this work. 
I am also grateful to Nakwoo Kim and Kostadinos Sfetsos 
for many useful discussions and comments on the manuscript, 
and Wolfgang Lerche, Jongwon Park, Jaemo Park, Piljin Yi 
for discussions. This work was partly supported by the CNNC. 


\newpage

\rnc{\baselinestretch}{1.20}

\end{document}